\documentclass[%
 reprint,
superscriptaddress,
 amsmath,amssymb,
 aps,
pra,
]{revtex4-1}

\usepackage{graphicx}% Include figure files
\usepackage{dcolumn}% Align table columns on decimal point
\usepackage{bm}% bold math
\usepackage{braket}
\usepackage{color}
\usepackage{subcaption}
\usepackage[font=footnotesize,labelfont=bf]{caption}
\usepackage[font=footnotesize,labelfont=bf]{subcaption}
\usepackage{hyperref}% add hypertext capabilities
\begin{document}

\preprint{APS/123-QED}

\title{Extreme spin squeezing in the steady state of a generalized Dicke model}% Force line breaks with \\

\author{Stuart J. Masson}\email{smas176@aucklanduni.ac.nz} 
\affiliation{Dodd-Walls Centre for Photonic and Quantum Technologies, New Zealand}
\affiliation{Department of Physics, University of Auckland, Auckland, New Zealand}

\author{Scott Parkins}\email{s.parkins@auckland.ac.nz}
\affiliation{Dodd-Walls Centre for Photonic and Quantum Technologies, New Zealand}
\affiliation{Department of Physics, University of Auckland, Auckland, New Zealand}

\date{\today}

\begin{abstract}
We present a scheme to generate steady-state atomic spin squeezing in a cavity QED system using cavity-mediated Raman transitions to engineer effective atom-photon  interactions, which include both linear and nonlinear (dispersive) atom-cavity couplings, on a potentially equal footing. We focus on a regime where the dispersive coupling is very large and find that the steady state of the system can in fact be a strongly spin-squeezed Dicke state, $\ket{N/2,0}$, of the atomic ensemble. These states offer Heisenberg-limited metrological properties and feature genuine multipartite entanglement among the entire atomic ensemble.
\end{abstract}

\maketitle

%%%%%%%%%%%%%%%%%%%%%%%%%%%%%%%%%%%%%%%%%%%%%%%%%%
%%%%%%%%%%%%%%%%%%%%%%%%%%%%%%%%%%%%%%%%%%%%%%%%%%
\section{Introduction}

Atom interferometers are useful for making precision measurements of acceleration, time, rotation, and, potentially, even gravitational waves \cite{TinoBook}. Interferometers employing uncorrelated states of $N$ atoms have a variance limited by the standard quantum limit (SQL), which scales like $1/N$, but suitably entangled atomic states could potentially reach the Heisenberg limit, where scaling like $1/N^2$ becomes the lower bound for the variance \cite{Bollinger96,Toth14}. Spin squeezed states \cite{Kitagawa93, Wineland92, Wineland94, Ma11,Gross12,Pezze18} are a popular choice to try to make measurements that are below the SQL and potentially approach the Heisenberg limit. Successful spin squeezing experiments have been carried out using atomic collisions in Bose-Einstein condensates \cite{Gross10,Riedel10,Bookjans11PRL1,Hamley12,Ockeloen13,Muessel14,Muessel15,Hoang16PNAS,Linnemann16,Kruse16}, quantum non-demolition measurements \cite{LouchetChauvet10,SchleierSmith10PRA,Leroux10PRL1,Chen11,Sewell12,Behbood13,Hosten16,Engelsen17}, and various other methods \cite{Hald99,Cox14,Bohnet14,Auccaise15,Cox16}. The best of these experiments have exceeded the SQL by 100-fold \cite{Hosten16}, but, for the large numbers of atoms that were involved, this is still nowhere near the corresponding Heisenberg limit. If this limit could be approached, then the actual number of atoms required for significant gains in precision could in fact be relatively small. 

A class of idealized states that can potentially reach the Heisenberg limit are the so-called Dicke states \cite{Holland93,Krischek11}, which are simultaneous eigenstates of the collective angular momentum operators $\hat{\mathbf{S}}^2$ and $\hat{S}_z$, denoted by $\ket{S,m}$. Here we will consider only symmetric Dicke states, for which the wave function is symmetric under particle exchange. 

Dicke states do not lend themselves well to characterization by conventional spin squeezing measures, which generally rely on a well-defined polarization of the spin state. Instead, to characterize the squeezing we consider the Dicke squeezing parameter \cite{Zhang14},
\begin{equation}
\xi_D = N\frac{(\Delta \hat{S}_z)^2 + 1/4}{\braket{\hat{S}_x^2+\hat{S}_y^2}} \, .
\end{equation}
This parameter gives us access to the metrological sensitivity relative to the SQL. For Mach-Zehnder interferometry, the variance is bounded by $\beta(\xi_D/N)$, where $\beta$ is a factor of order one \cite{Zhang14}. The parameter also provides a lower bound for the entanglement depth of $\lceil \xi_D^{-1} -2 \rceil$ \cite{Duan11}, where $\lceil x\rceil$ denotes the minimum integer no less than $x$. Considering this bound, we can see that the limit for entanglement, and thus metrological gain, is $\xi_D = 1/2$.

For a Dicke state $\ket{N/2,m}$, the Dicke squeezing parameter is given by
\begin{equation}
\xi_D = \frac{1}{N+2 - \frac{4m^2}{N}}\label{limit} \, .
\end{equation}
This means that the Dicke states offer near Heisenberg limited metrological sensitivity for $N\gg \{2,m\}$. In addition, for $m < \sqrt{N}/2$ the entanglement depth must be the size of the entire atomic ensemble. There have been proposals for schemes to prepare Dicke states of an atomic ensemble (see, e.g., \cite{Raghavan01,Duan03,Stockton04,Kiesel07,Thiel07,Prevedel09,Wieczorek09,Haas14}), typically based upon conditional or probabilistic processes, and possibly also feedback of some sort. Alternatively, schemes using collisions in spin-1 Bose-Einstein condensates are possible and have been implemented \cite{Lucke14,Luo17,Zou18}.

Here we propose an approach that can, in principle, prepare an arbitrary Dicke state as the {\em steady state} of a cavity QED system with a suitably engineered atom-cavity interaction. In particular, this interaction is described by the so-called ``generalized Dicke model'', which may be engineered, in an optical cavity QED setting with alkali atoms, via laser- and cavity-driven Raman transitions between ground-state electronic sublevels of the atoms \cite{Dimer07,Grimsmo13PRA,Grimsmo13JPB,Zhiqiang17,Masson17,Zhiqiang18}. 

We describe the model and the specific parameter regime in which the desired steady-state behavior - a near pure Dicke state - is obtained, and explaining the mechanism that allows that steady state to exist. We then explain how that same mechanism means that reaching the steady state through natural evolution is an extremely slow process, which leads us to consider modified schemes. The first of these produces the state heralded by the detection of a single photon in the output channel, while the second makes use of time variation of one of the parameters of the model. These methods allow for the preparation of the state on much faster timescales. We conclude with a discussion of a specific realization in an optical cavity QED system and show that significant squeezing and high fidelity states can be achieved with feasible experimental conditions.

%%%%%%%%%%%%%%%%%%%%%%%%%%%%%%%%%%%%%%%%%%%%%%%%%%
%%%%%%%%%%%%%%%%%%%%%%%%%%%%%%%%%%%%%%%%%%%%%%%%%%
\section{System and model}

We consider a generalized Dicke model for $N$ two-level atoms and a single mode of the electromagnetic field, as described by the master equation \cite{Dimer07,Bhaseen12,Grimsmo13PRA,Grimsmo13JPB,Zhiqiang18,Kirton18}
\begin{equation}
\dot{\rho} = -i[\hat{H},\rho] + \kappa \mathcal{D}[\hat{a}]\rho,\label{dickestatemasterequation}
\end{equation}
with the Hamiltonian
\begin{equation}
\hat{H} = \omega_0 \hat{S}_z + \omega \hat{a}^\dagger \hat{a} + \frac{\lambda}{\sqrt{N}}(\hat{a}+\hat{a}^\dagger)(\hat{S}_++\hat{S}_-) + \frac{U}{N} \hat{S}_z \hat{a}^\dagger \hat{a} ,\label{dickestateprepdickemodel}
\end{equation}
where $\mathcal{D}[\hat{a}]\rho$ represents the superoperator
\begin{equation}
\mathcal{D}[\hat{a}]\rho = 2\hat{a} \rho \hat{a}^\dagger -  \hat{a}^\dagger \hat{a} \rho - \rho \hat{a}^\dagger \hat{a}.
\end{equation}
Here $\hat{a}$ ($\hat{a}^\dagger$) is the annihilation (creation) operator for the quantized cavity mode, and $\hat{S}_i$ are collective atomic spin operators satisfying the usual angular momentum commutation relations. The linear atom-field coupling strength is denoted by $\lambda$, the nonlinear (or dispersive) coupling strength is given by $U$, $\omega$ and $\omega_0$ are the cavity and atomic resonance frequencies, respectively, and $\kappa$ is the cavity field decay rate.

This system possesses a rich phase diagram, which has been studied both semi-classically \cite{Keeling10,Bhaseen12} and quantum mechanically \cite{Grimsmo13JPB}. In \cite{Grimsmo13JPB}, it was noticed that as $U$ becomes very large compared to the other parameters, the atomic state can become strongly squeezed in the $\hat{S}_z$ spin component. The present work looks more closely at, and provides an explanation for, this large-$U$ behavior. 

Before continuing, we note that our Hamiltonian is similar in structure to that put forward in \cite{Luo12}, where a method based on time-varying parameters was proposed for preparing Dicke states of donor nuclear spins in silicon. The discrete stepping method we describe in Section~\ref{timedep} can be viewed as parallel to the method in \cite{Luo12}.

%%%%%%%%%%%%%%%%%%%%%%%%%%%%%%%%%%%%%%%%%%%%%%%%%%
%%%%%%%%%%%%%%%%%%%%%%%%%%%%%%%%%%%%%%%%%%%%%%%%%%
\section{Steady state behaviour}

%%%%%%%%%%%%%%%%%%%%%%%%%%%%%%%%%%%%%%%%%%%%%%%%%%
\subsection{Expectation values}

\begin{figure}[t!]
\includegraphics[width=0.5\textwidth]{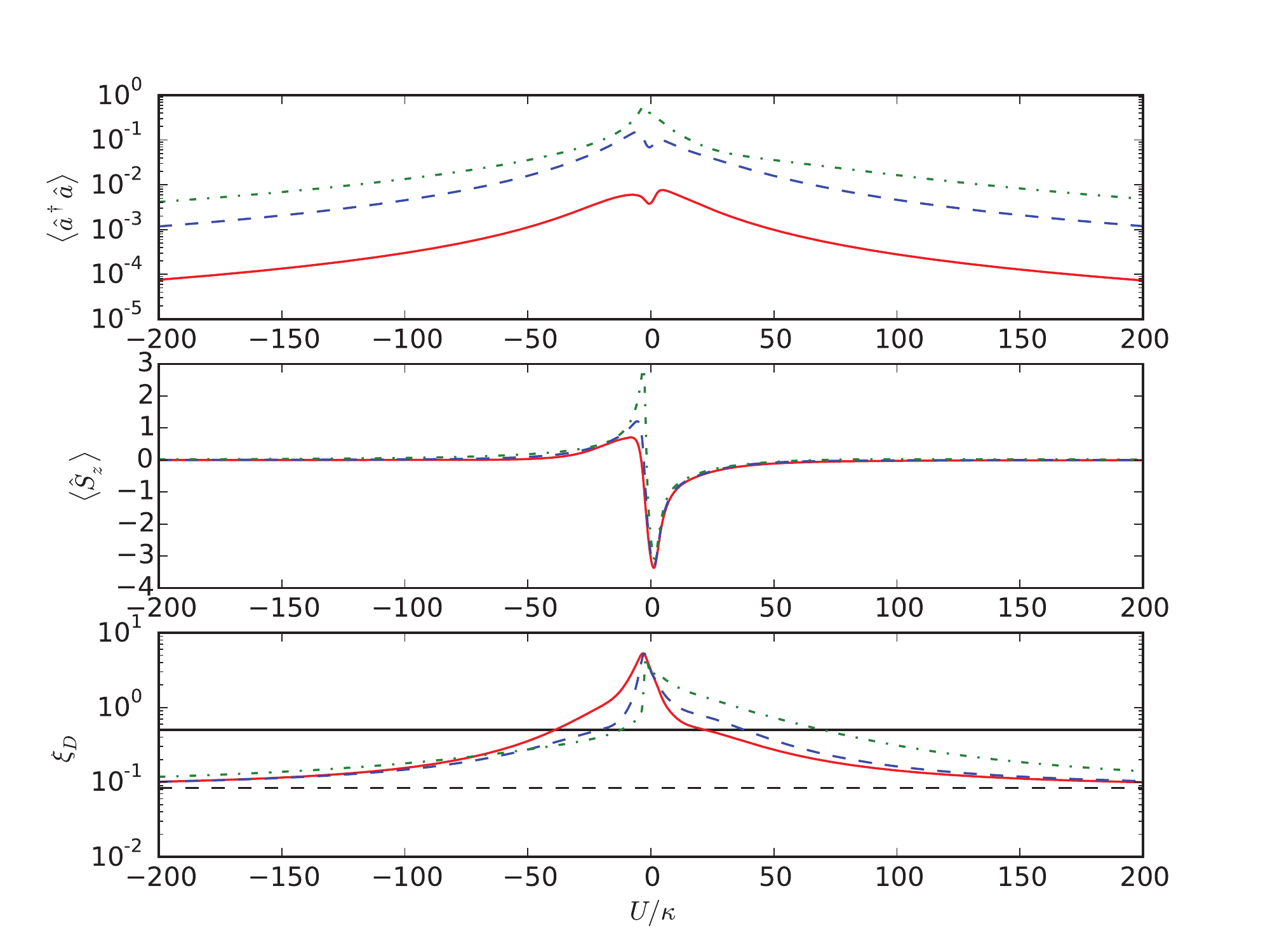}
\caption{Steady state expectation values for $\set{\omega,\omega_0}/\kappa=\set{1.0,0.2}$ and $N=10$, with $\lambda/\kappa=0.05$ (red solid line), $\lambda/\kappa=0.2$ (blue dashed) and $\lambda/\kappa=0.4$ (green dash-dotted). The black lines in the plot of $\xi_D$ are the standard quantum limit (solid) and the ideal limit of $1/(N+2)=1/12$ (dashed).\label{varyU}}
\end{figure}

We consider the steady state behavior of the master equation (\ref{dickestatemasterequation}) as we vary $U$. Fig.~\ref{varyU} displays various properties of the system as $U$ is varied for several different values of (linear) coupling strength $\lambda$. In particular, it plots the steady-state values of the mean intracavity photon number, $\braket{\hat{a}^\dag \hat{a}}$, the collective atomic inversion, $\braket{\hat{S}_z}$, and the Dicke squeezing parameter, $\xi_D$.

\begin{figure}[b!]
\includegraphics[width=0.495\textwidth]{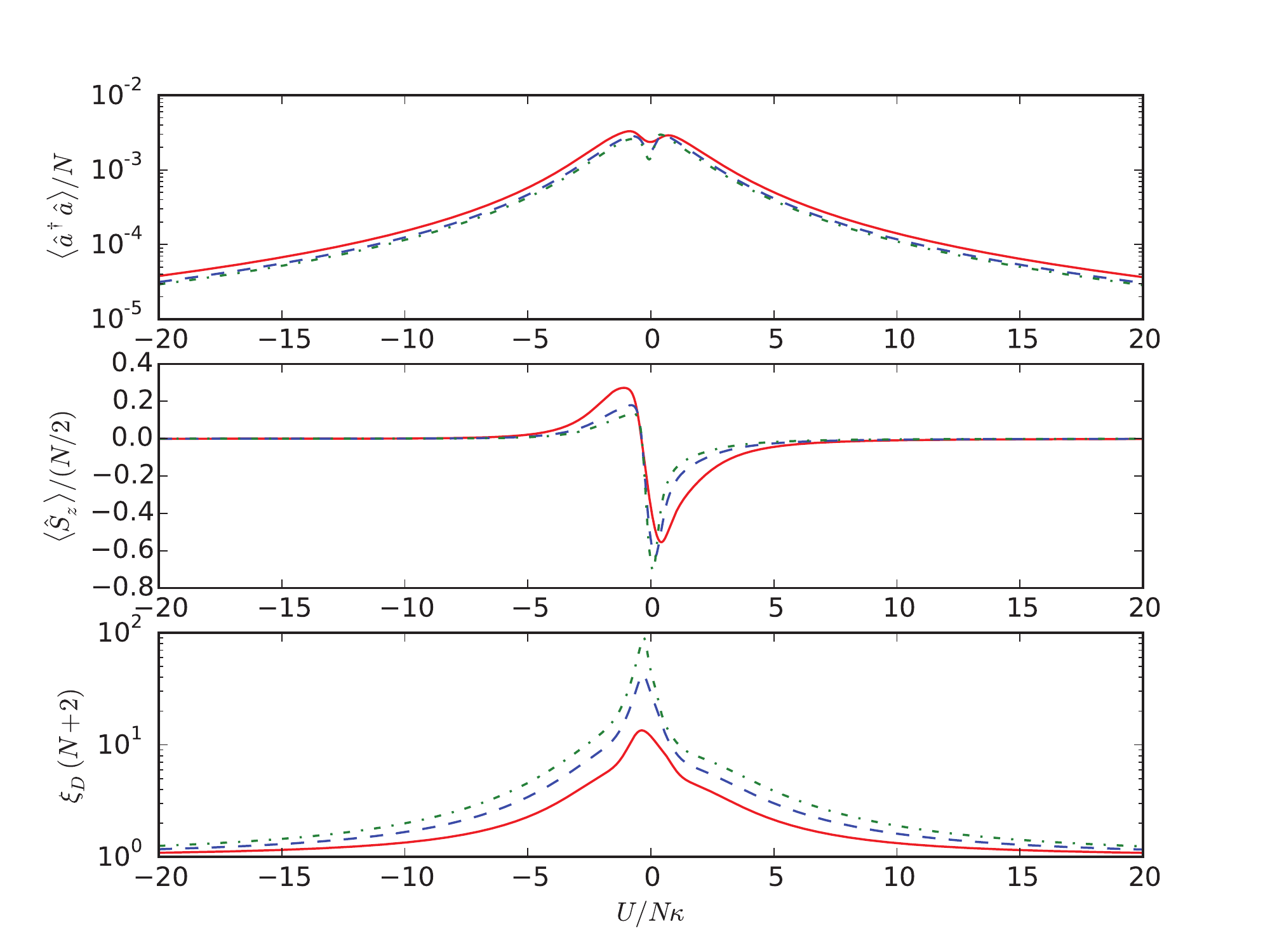}
\caption{Steady state expectation values for $\set{\omega,\omega_0,\lambda}/\kappa=\set{1.0,0.2,0.1}$ with $N=4$ (red solid line), $N=8$ (blue dashed), and $N=12$ (green dash-dotted). Here the Dicke squeezing parameter, $\xi_D$, is given in proportion to the ideal limit, $1/(N+2)$.\label{varyUvN}}
\end{figure}

For small $|U|$ we see that the properties of the system depend rather sensitively on $U$ and $\lambda$. However, for large $|U|$ we observe a simpler, monotonic dependence: the mean photon number decreases steadily, becoming very small, the atomic inversion converges quite rapidly to zero, and the Dicke squeezing parameter approaches the value $1/(N+2)$. This indicates that the system settles predominantly into the Dicke state $\ket{N/2,0}$, corresponding to genuine multipartite entanglement of the entire ensemble and Heisenberg-limited metrological sensitivity. The relative lack of sensitivity to the coupling strength $\lambda$ is in direct contrast to the traditional and well-known superradiant behavior of the Dicke model that occurs when $U$ is small or negligible \cite{Dimer07,Keeling10,Bhaseen12,Grimsmo13PRA,Grimsmo13JPB,Zhiqiang17,Shammah17,Zhiqiang18,Kirton18}.

In Fig.~\ref{varyUvN} we plot $\braket{\hat{a}^\dag \hat{a}}$, $\braket{\hat{S}_z}$, and $\xi_D$ as a function of $U/N$ for several different values of $N$ and observe that the same general pattern holds. We do note, though, that for a given (large) value of $U/N$ the Dicke squeezing parameter is closer to its ideal limit of $1/(N+2)$ for smaller $N$, though it should be noted that the absolute squeezing is still larger for higher $N$.

%%%%%%%%%%%%%%%%%%%%%%%%%%%%%%%%%%%%%%%%%%%%%%%%%%
\subsection{Dicke state preparation}

\subsubsection{$\ket{N/2,0}$}

\begin{figure}[b!]
\includegraphics[width=0.5\textwidth]{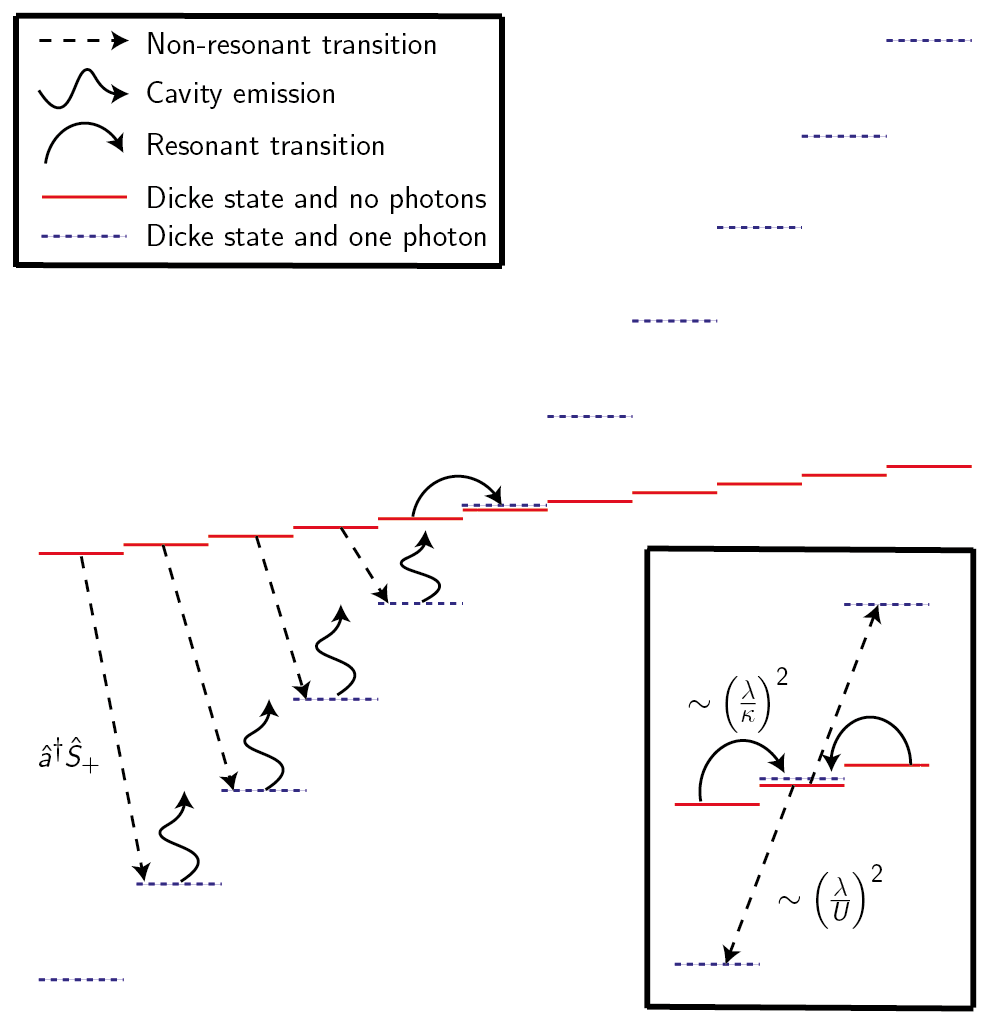}
\caption{Approximate level diagram for large positive $U \gg \omega, \omega_0, \lambda$. The dominant evolution pathway is illustrated.
\label{energylevels}}
\end{figure}

If we consider the energy level structure of the system in the limit of large $|U|$ (see Fig.~\ref{energylevels}), and consider the possible transitions between states as allowed by the atom-cavity coupling Hamiltonian and by cavity photon emission, then it is possible to understand why the steady atomic spin state $\ket{N/2,0}$ emerges. Let us use the notation $\ket{N/2,m,n} = \ket{N/2,m} \otimes \ket{n}$, where $\ket{n}$ is the $n$-photon Fock state of the cavity mode. 

Now, consider the state $\ket{N/2,m,0}$ with $m<-1$. The only processes by which a transition from this state may occur are described by the terms $\hat{a}^\dag \hat{S}_+$ and $\hat{a}^\dag \hat{S}_-$ in the Hamiltonian. The corresponding transitions are $\ket{N/2,m,0} \rightarrow \ket{N/2,m+1,1}$ and $\ket{N/2,m,0} \rightarrow \ket{N/2,m-1,1}$, respectively. 
For large $U$ (i.e., $U$ much larger than any of the other parameters), both of these transitions are off-resonant by $\sim U(m\pm 1)/N$. Whilst both transitions are strongly off-resonant, for negative $m$, the former is less so by an amount $\sim 2U/N$, and will therefore be favoured, causing a net evolution towards states of larger $m$ (in combination with cavity photon emissions, which cause the transitions $\ket{N/2,m+1,1} \rightarrow \ket{N/2,m+1,0}$). Similarly, for $m>1$, transitions via $\hat{a}^\dag \hat{S}_-$ will be preferred, causing a net evolution towards states of smaller $m$. Hence, whatever its initial state, the system will evolve towards the center of the spin angular momentum ladder. 

On reaching the states $\ket{N/2,\pm 1,0}$, ``inward'' transitions to the state $\ket{N/2,0,1}$ become approximately resonant (for $\lambda\sim\omega\sim\omega_0$), and, following emission of the cavity photon, the state $\ket{N/2,0,0}$ is prepared. Importantly, provided $U/N \gg \{\omega_0,\omega,\lambda,\kappa\}$, transitions out of the state $\ket{N/2,0,0}$ to $\ket{N/2,\pm 1,1}$ (followed by photon emission to $\ket{N/2,\pm 1,0}$) will be much weaker, due to the much larger energy gap ($\sim U/N$), than the inward transitions. Hence, the system essentially becomes ``trapped'' in the Dicke state $\ket{N/2,0,0}$. 

\subsubsection{$\ket{N/2,m}$} 

\begin{figure}[b!]
\includegraphics[width=0.5\textwidth,]{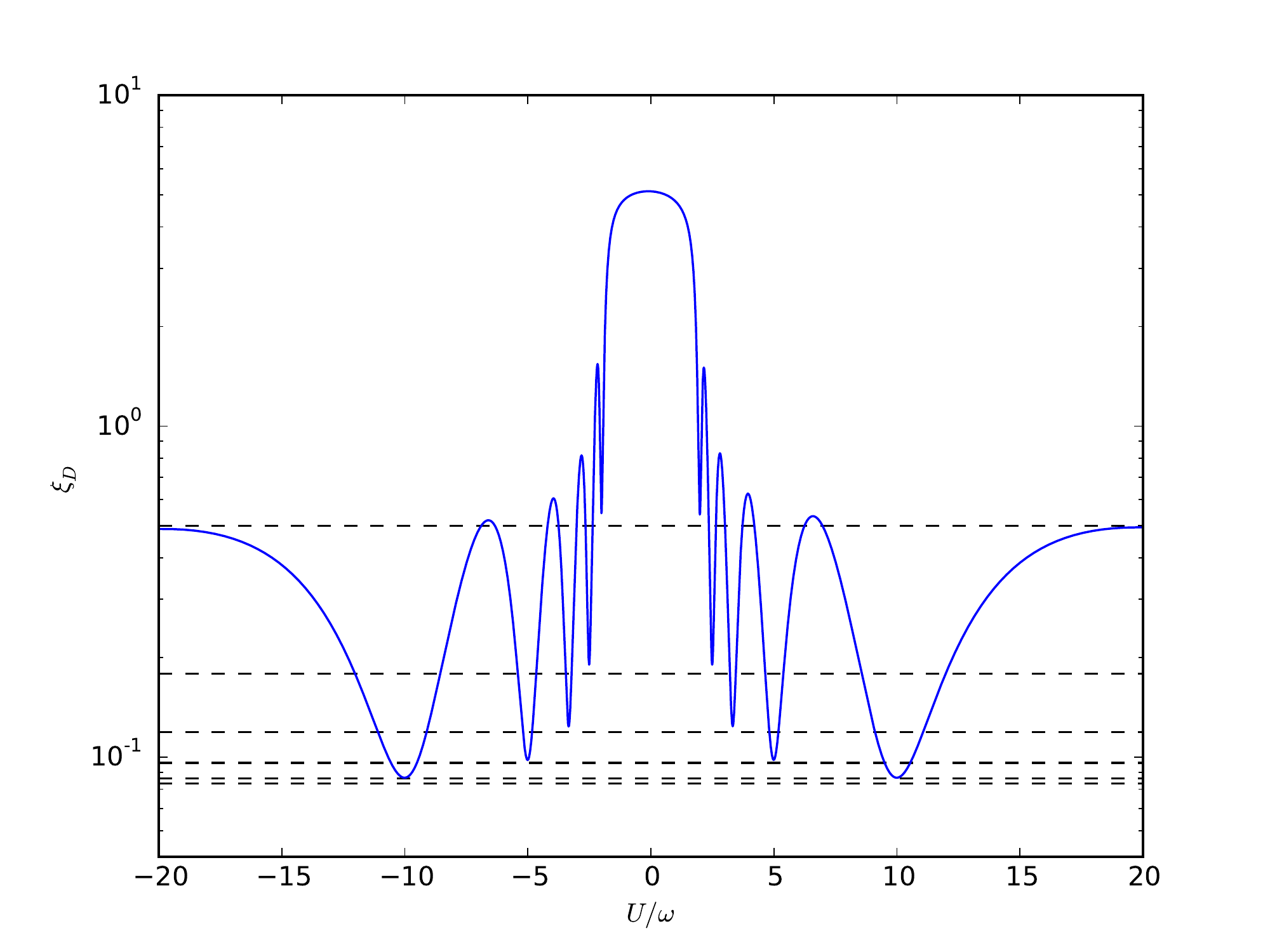}
\caption{Steady state Dicke squeezing parameter $\xi_D$ as a function of $U/\omega$ for $\set{\omega,\omega_0,\lambda}/\kappa=\set{100.0,0.2,0.2}$, with $N=10$. The dashed lines are those predicted by Eq.~(\ref{limit}).
\label{varyUarbd}}
\end{figure}

If we rewrite the Hamiltonian as
\begin{equation}
\hat{H} = \omega_0 \hat{S}_z + \hat{a}^\dagger \hat{a} \left( \omega + \frac{U}{N}\hat{S}_z \right) + \frac{\lambda}{\sqrt{N}}(\hat{a}+\hat{a}^\dagger)(\hat{S}_++\hat{S}_-) ,
\end{equation}
then we see that for $U = -\omega N/m$ (with $\omega \gg \kappa$) the states $\ket{N/2,m,n}$ are degenerate for all $n$. This means that the transitions $\ket{N/2,m\pm1,0}\rightarrow\ket{N/2,m,1}$ are now the resonant transitions. Following the same arguments as given above, one finds that this shifts the trapped state to $\ket{N/2,m,0}$. Hence, by tuning parameters, the steady state can be adjusted to an arbitrary Dicke state. Fig.~\ref{varyUarbd} illlustrates this possibility in the form of Dicke squeezing ``resonances'' occurring at $U/\omega =-N/m$ ($m=\pm 1,\ldots ,\pm 5$) for $N=10$.

%%%%%%%%%%%%%%%%%%%%%%%%%%%%%%%%%%%%%%%%%%%%%%%%%%
%%%%%%%%%%%%%%%%%%%%%%%%%%%%%%%%%%%%%%%%%%%%%%%%%%
\section{Dynamic behaviour}

A problem with using the above approach to produce Dicke states is the timescale involved with reaching the steady state. To estimate the timescale of some transition $\ket{N/2,m,0} \rightarrow \ket{N/2,m+1,0}$, we consider an analytic quantum trajectory approach with a state
\begin{align}
\notag\ket{\psi(t)} = \alpha(t) \ket{N/2,m,0} & + \beta(t) \ket{N/2,m-1,1} \\&+ \gamma(t) \ket{N/2,m+1,1} , \label{dickestatepreptrajectorystate}
\end{align}
where $\alpha(0)=1$, $\beta(0)=\gamma(0)=0$, and we assume that no more than one photon is present in the cavity mode. We then calculate the Schr\"{o}dinger evolution of the state with an effective Hamiltonian
\begin{equation}
\hat{H}_{\mathrm{eff}} = \hat{H} - i\kappa \hat{a}^\dagger\hat{a}
\end{equation}
where $\hat{H}$ is the generalized Dicke model, (\ref{dickestateprepdickemodel}). The cavity output flux is proportional to $|\beta(t)|^2 + |\gamma(t)|^2$, and so, taking $U \gg \{\omega,\omega_0,\lambda,\kappa\}$, the timescale for the transition $\ket{N/2,m,0} \rightarrow \ket{N/2,m\pm1,0}$ (except for $m=\pm1$)  can be calculated approximately as
\begin{equation}
T_{m} \approx \frac{U^2}{2N\kappa \lambda^2} \frac{(m\pm1)^2}{N/2(N/2+1) - m(m\pm1)}.
\end{equation}
The total time will thus scale as $U^2$ and so be extremely large in the parameter regime considered. For constant $U/N$, more atoms increases the time due to requiring more steps to reach $m=0$.

Evolution from the fully polarized states $\ket{N/2,\pm m}$ requires the entire population to undergo every single step to $\ket{N/2,0}$. However, we could instead use an initial coherent spin state (CSS) \cite{Arecchi72},
\begin{equation}
\ket{\eta} = (1+|\eta|^2)^{-j} \sum\limits_{m=-j}^{j} \begin{pmatrix} N \\ j+m \end{pmatrix}^{1/2} \eta^{j+m} \ket{j,m},
\end{equation}
where $\eta = \mathrm{e}^{-\mathrm{i}\varphi} \tan\left(\theta/2\right)$ and $\{\theta,\varphi\}$ are spherical coordinates. With a suitable choice of $\eta$ (e.g., $\eta=1$), this improves the evolution time because much of the population in this state overlaps with or is close to the steady state, while very little of the state is in or near the fully polarized end states. Such an initial state also immediately shows Dicke squeezing upon evolution, while the fully polarized initial state first evolves to an anti-squeezed state before slowly approaching the squeezed steady state.

While the initial CSS improves the short term generation of Dicke squeezing, the time for significant squeezing to appear is still very large even for very small ensembles.  Remembering that increased $N$ or $U$ dramatically increases this time, even with an optimized initial state, the evolution to the steady state is likely to be prohibitively slow.

%%%%%%%%%%%%%%%%%%%%%%%%%%%%%%%%%%%%%%%%%%%%%%%%%%
%%%%%%%%%%%%%%%%%%%%%%%%%%%%%%%%%%%%%%%%%%%%%%%%%%
\section{Probabilistic preparation}

\begin{table}[b!]
\begin{center}
\begin{tabular}{|c|c|c|c|c|c|}
\hline
$N$ & Efficiency & Fidelity & $\bar{\xi}_D$ ($E_D$) & Within 1\% & Within 10\% \\ \hline
10 & 40.3\% & 99.94\% & 0.0839 (10) & 91.8\% & 98.7\% \\ \hline
20 & 31.8\% & 99.91\% & 0.0460 (20) & 81.3\% & 97.6\% \\ \hline
50 & 20.8\% & 99.89\% & 0.0197 (49) & 42.0\% & 95.2\% \\ \hline
100 & 15.5\% & 99.52\% & 0.0115 (85) & 0.0\% & 89.2\% \\ \hline
\end{tabular}
\caption{Properties of 5000 trajectories (1000 for $N=100$) with $\set{\omega,\omega_0,\lambda,U/N}/\kappa=\set{1.0,0.2,0.1,100}$. All properties except the efficiency are only for successful trajectories, i.e. trajectories that have exactly one jump before $\kappa t = 52$. Fidelity is the average fidelity with the Dicke state $\ket{N/2,0}$, $\bar{\xi}_D$ is the average Dicke squeezing, $E_D$ is the minimum entanglement depth and within 1\% (10\%) is the percentage of trajectories with squeezing within 1\% (10\%) of the ideal limit for Dicke squeezing.\label{popprep}}
\end{center}
\end{table}

One potential method for preparing the steady state in a shorter time span involves using an initial CSS for the atomic ensemble and probabilistic photon detection. A photon detection collapses the state into a superposition weighted by how likely the states in the initial superposition were to have produced a photon. The resultant superposition is dominated by the resonant state, as its neighbors are by far the most likely states to produce photons. Single photon detection can thus produce extremely high fidelity Dicke states in very short time frames. The production of highly non-classical states via heralded single photon detection schemes has been proposed and implemented in other systems \cite{Chia08,Christensen13,Casabone13,Mcconnell13,Mcconnell15,Christensen14,Chen15,Welte17,Davis18PRL}.

The probability of creating the steady state $\ket{N/2,m,0}$ with this method is the sum of the populations in the states $\ket{N/2,m\pm1,0}$. For example, if the desired state is $\ket{N/2,0,0}$ then a CSS with $\eta=1$ maximizes the overlap and the probability of success is
\begin{align}
P &= 2^{-N} \left[\begin{pmatrix} N \\ \frac{N}{2} +1 \end{pmatrix} + \begin{pmatrix} N \\ \frac{N}{2} -1 \end{pmatrix} \right].\label{probsuccess}
\end{align}
For small atomic ensembles this proves to be a fairly efficient method. The results of trajectory simulations are shown in Table \ref{popprep}. Here we see reasonably high success rates that decline with $N$, and match reasonably with those predicted in (\ref{probsuccess}) (e.g., for $N=10 (100)$ the prediction is 41.3\% (15.6\%)). The fidelity is extremely high, and, since the rate of these resonant transitions is independent of $U$, can be made arbitrarily higher without increasing the time taken or reducing the success probability. We also see very strong squeezing and high levels of multipartite entanglement. For $N=100$, the squeezing has metrological sensitivity -19.39dB improved over the SQL. This means that this method can produce competitive levels of squeezing with a relatively high success probability in a very short time span for much smaller numbers of atoms than usual spin squeezing techniques. 

However, creating significantly better squeezing comes at a cost. To match the $\sim 3000$ depth multipartite entanglement in Ref. \cite{Mcconnell15} would require a Dicke state $\ket{1500,0}$ with an associated success rate of 2.9\%, though we note much stronger entanglement could be created. With $10^6$ atoms the probability has become a fractional 0.16\%. A potentially useful fact is that the speed at which the transition occurs scales as $1/N$ for central Dicke states. If the CSS could be recreated very quickly then it might be possible to do more iterations with higher $N$, and thus help to account for some of the inefficiency. Alternatively, a spin squeezed state aligned along the equator of the collective Bloch sphere will have enhanced population in the central Dicke states, including $\ket{N/2,\pm1}$, and so the use of a spin squeezed initial state could enhance the efficiency of this probabilistic scheme.

There is a special case for this method for which the efficiency would be unity. If parameters are set such that the resonant transition is between $\ket{N/2,\pm N/2,0}\rightarrow\ket{N/2,\pm N/2 \mp 1,1}$ then preparing the initial state in $\ket{N/2,\pm N/2}$ would always mean a jump within a short time frame. These so-called $W$ states \cite{Dur00} (symmetric Dicke states with a single excitation) have numerous applications in quantum information \cite{Shi02,Agrawal06,Zheng06} and here we have a method that can produce them with close to unit efficiency on a short time scale and in the steady state. The error in the fidelity can be approximated by expanding the space of the state given in (\ref{dickestatepreptrajectorystate}) to include two photons. This gives an error of
\begin{equation}
\varepsilon \approx \frac{15\lambda^2N^2}{4U^2}.
\end{equation}
We can see that the error is kept approximately constant for constant $U/N$, and, since the transition time here is independent of $U$ and $N$, this method scales very well to larger atomic ensembles.

\begin{figure}[b!]
\includegraphics[width=0.45\textwidth]{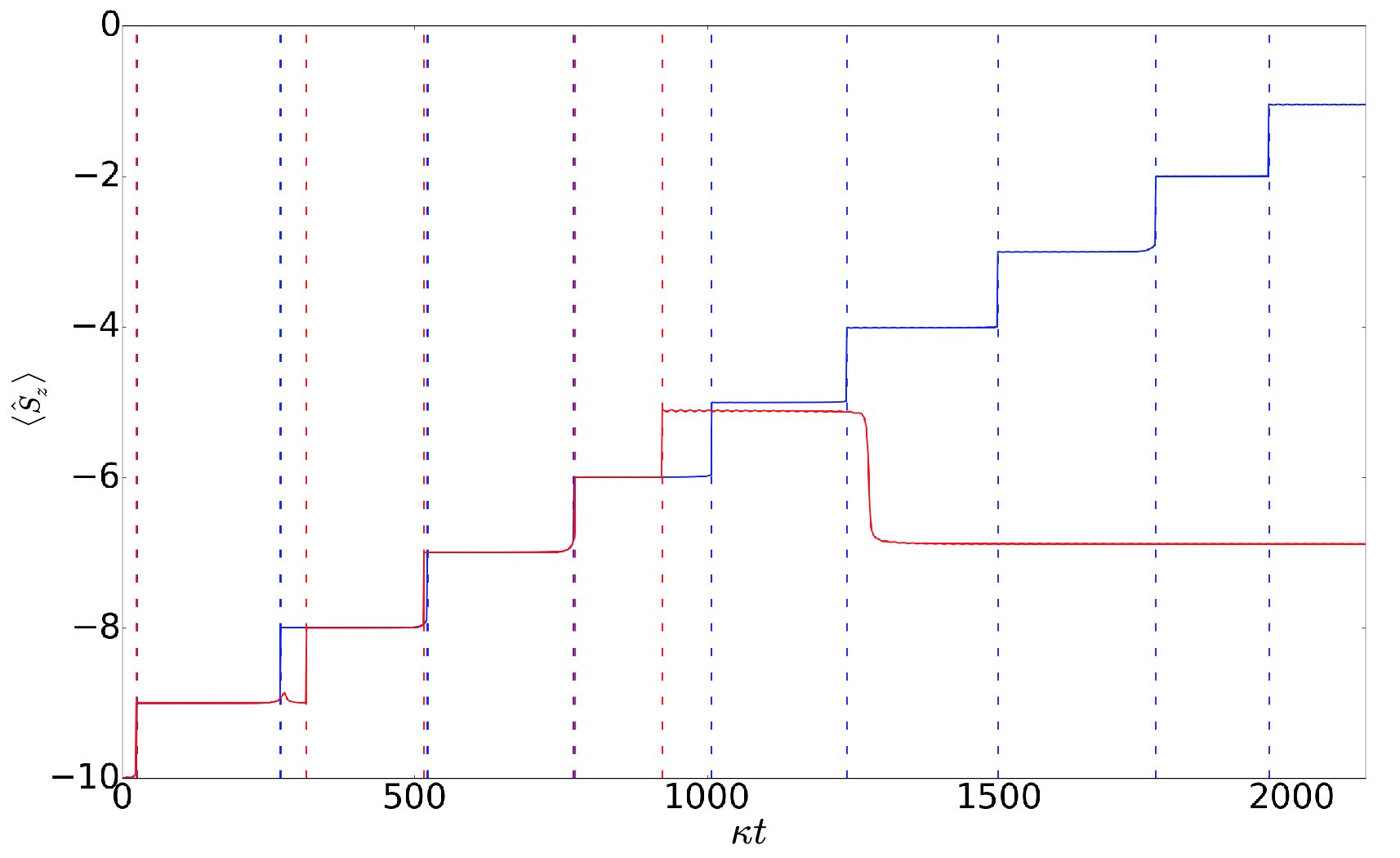}
\caption{The time evolution of $\braket{\hat{S}_z}$ with photon detections (vertical lines) for a successful trajectory (blue) and a failure (red), parameters of $\set{\omega_0,\lambda,U}/\kappa = \set{0.2,0.2,1000.0}$, $N=20$ and $\omega$ linearly varied as $\omega(t)/\kappa = 470 - 0.2 \kappa t$.\label{photons}}
\end{figure}

%%%%%%%%%%%%%%%%%%%%%%%%%%%%%%%%%%%%%%%%%%%%%%%%%%
%%%%%%%%%%%%%%%%%%%%%%%%%%%%%%%%%%%%%%%%%%%%%%%%%%
\section{Time dependent parameters\label{timedep}}

We have shown that, by correct choice of parameters, the steady state can be tuned to a desired Dicke state. We have also shown that if the system is one Dicke state either side of the steady state, then the evolution to that state is very fast. Thus, with complete control of the parameters, it should be possible to step the system from $\ket{N/2,-N/2,0}\rightarrow\ket{N/2,-N/2+1,0}\rightarrow...\rightarrow\ket{N/2,0,0}$. This requires time dependence in $\omega$.

\begin{figure}[b!]
\begin{subfigure}[b]{0.45\textwidth}
\includegraphics[width=\textwidth]{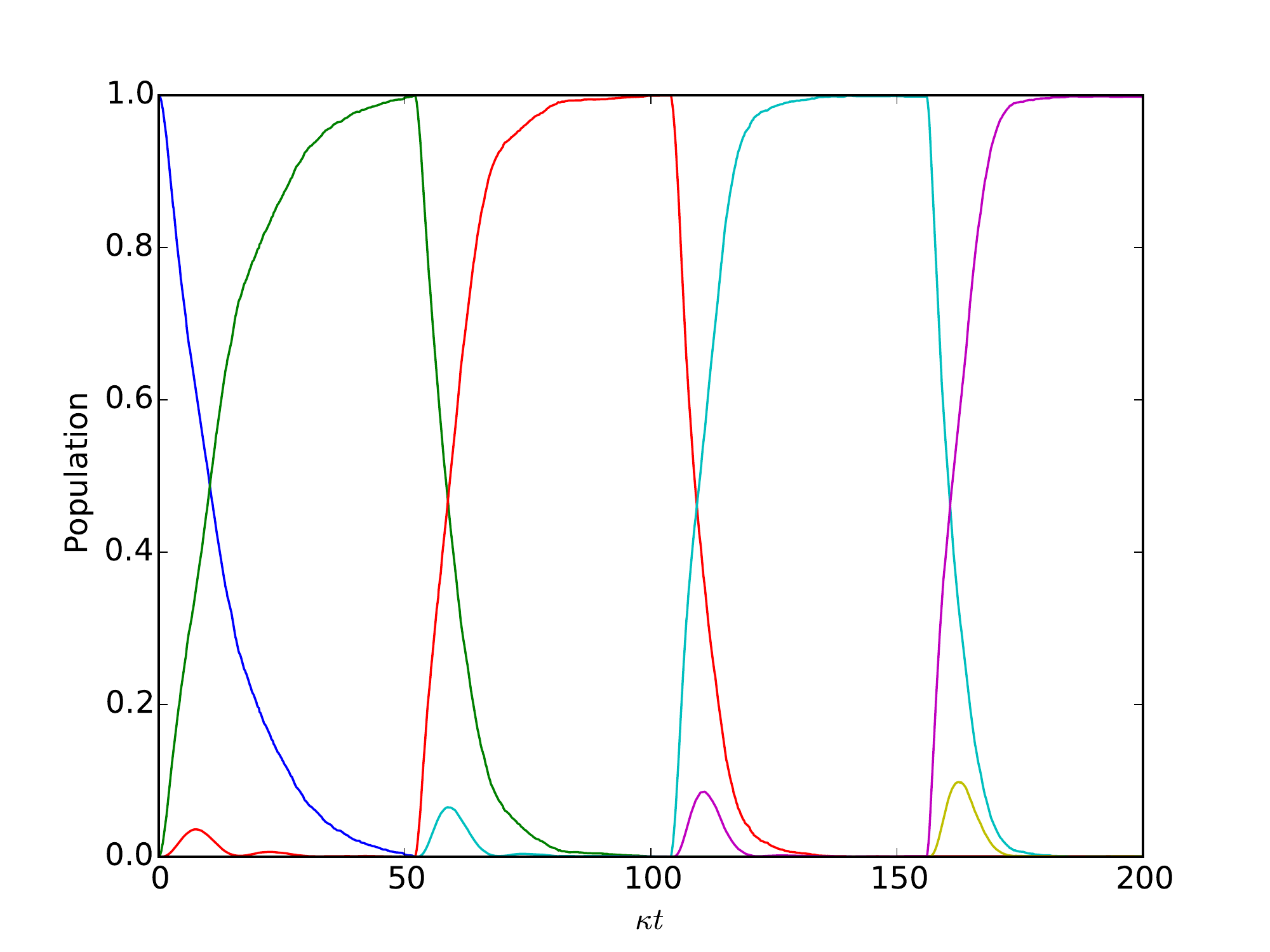}
\caption{Time evolution of Dicke state populations with blue, green, red, cyan and pink (or peaks from left to right) being $m = -5, -4, -3, -2, -1$ respectively.\label{popdsc}}
\end{subfigure}
\begin{subfigure}[b]{0.45\textwidth}
\includegraphics[width=\textwidth]{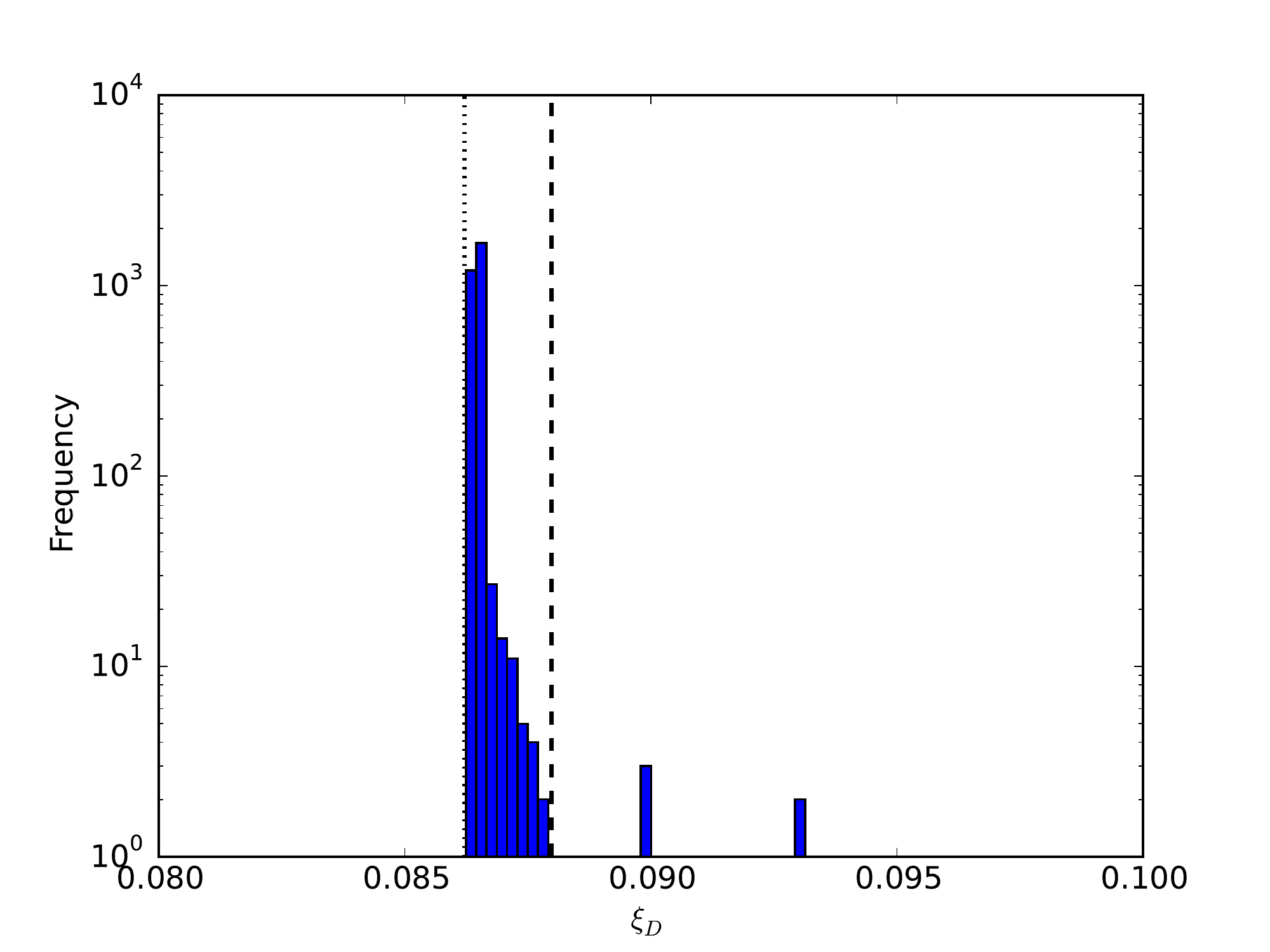}
\caption{Histogram of final squeezing. The vertical lines are the ideal limit (dotted) and steady state squeezing (dashed).\label{sqzdsc}}
\end{subfigure}
\caption{Properties of 985 successful trajectories out of 1000 total trajectories, with parameters of $\set{\omega_0,\lambda,U}/\kappa=\set{0.2,0.2,500.0}$, $N=10$ and time dependent $\omega$ discretely stepped such that it makes each transition from $\ket{0,-5}\rightarrow\ket{0,-1}$ resonant in turn. The system is held at each step for $\kappa t_h = 52$.\label{dsc}}
\end{figure}

Successful runs can be post-selected based on the number of photons emitted. If that number is equal to the total number of steps necessary then the trajectory is deemed successful, otherwise it is a failure. Experimentally, this would require an extremely efficient single photon detector. However, it is also possible to differentiate successes and failures by the pattern of the photon emissions. If the state is ``dragged'' all the way to the middle, then the photons should come at relatively frequent intervals throughout the length of the experiment. This means that if a photon is detected near the end of the trajectory then all the photons that came before can be inferred, as shown in Fig. \ref{photons}.

\begin{figure}[b!]
\includegraphics[width=.42\textwidth]{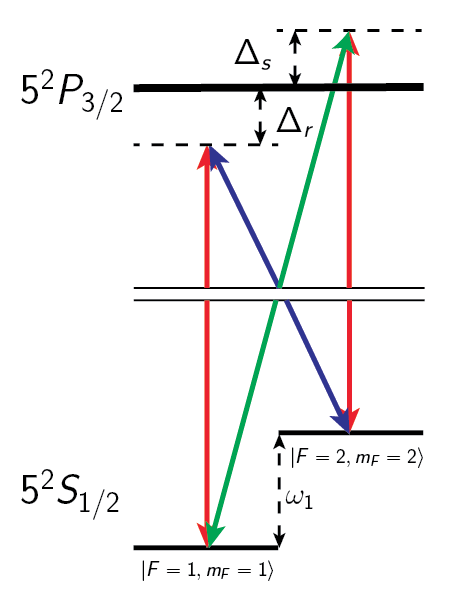}
\caption{Level scheme for the implementation of the generalized Dicke model with $^{87}$Rb atoms. Transitions are driven with Raman transitions composed of a cavity mode (solid red) and $\sigma_+$-polarized (dot-dashed green) and $\sigma_-$-polarized (dashed blue) laser fields. Raman transitions are detuned either side of the excited manifold to maximize the non-linear term. Note that the level diagram is not drawn to scale.}
\end{figure}

Linear variation of $\omega$, such that the system is initially resonant for creation of the $W$ state, and then ends at zero, such that the final resonant transition is to the central Dicke state, thus has some probability of producing that central Dicke state with very high fidelity. For ten atoms, an ensemble of quantum trajectories produces a $58.0\%$ success rate, a best squeezing of $\xi_D = 0.086$ and an average squeezing of $\bar{\xi}_D = 0.100$.

More successful, if perhaps more experimentally challenging, is discretely stepping $\omega$, i.e., implementing the time dependence
\begin{equation}
\omega(t) = -\frac{U}{N}\left(-\frac{N}{2}+j\right)\;\; \mathrm{for} \;\; t_h(j-1) \leq t < t_hj \, ,
\end{equation}
where $t_h$ is the time the system is held at each step and $j$ is an integer stepping from $1\rightarrow \Delta m$. We know that the slowest transition is the first one, and so we set $t_h$ such that the first transition will almost certainly have occurred.

The population transfer and squeezing for this discrete stepping approach with $N=10$ are shown in Fig.~\ref{dsc}. The best squeezing is $\xi_D^{\mathrm{min}}=0.086$, but with a greatly improved average squeezing of $\bar{\xi}_D = 0.087$. This means that the \emph{average} successful trajectory has near Heisenberg-limited metrological sensitivity. There are also much higher success rates, with 98.6\% of 1000 trajectories being successful. This scales well with $N$ as well, with a success rate of 98.5\% (83.9\%) for $N=20$ (50).

%%%%%%%%%%%%%%%%%%%%%%%%%%%%%%%%%%%%%%%%%%%%%%%%%%
%%%%%%%%%%%%%%%%%%%%%%%%%%%%%%%%%%%%%%%%%%%%%%%%%%
\section{Cavity-QED realisation}

%%%%%%%%%%%%%%%%%%%%%%%%%

\subsection{Microscopic parameters}
%%%%%%%%%%%%%%%%%%%%%%%%%
We now consider the optical cavity-QED realization of the Dicke model described in \cite{Dimer07,Grimsmo13JPB,Zhiqiang18}. The necessary Hamiltonian is produced via resonant Raman transitions in a dilute ensemble of $^{87}$Rb atoms interacting with a high finesse optical cavity mode. Here the effective parameters are given in terms of the microscopic parameters by \cite{Zhiqiang18}
\begin{align}
\notag \omega_0 &= \omega_1 - \frac{1}{2}\left(\omega_s - \omega_r\right) \\
& + \frac{1}{6} \left( \frac{\Omega_r^2}{\Delta_r} - \frac{\Omega_r^2}{\Delta_r - \omega_1} - \frac{\Omega_s^2}{\Delta_s} + \frac{\Omega_s^2}{\Delta_s + \omega_1} \right) , \\
\omega &= \omega_c - \frac{1}{2}\left(\omega_r + \omega_s\right) + \frac{N}{3}\left(\frac{g^2}{\Delta_s} + \frac{g^2}{\Delta_r}\right) , \\
U &= \frac{2 N}{3} \left(\frac{g^2}{\Delta_s} - \frac{g^2}{\Delta_r}\right) , \\
\lambda_{r,s} &= \frac{\sqrt{3N}g\Omega_{r,s}}{12\Delta_{r,s}} ,
\end{align}
where $\{\lambda_r,\lambda_s\}$ are the Raman coupling strengths for the rotating and counter-rotating terms respectively, and $\omega$ and $\omega_0$ are the \emph{effective} cavity and atomic frequencies respectively, defined by combinations of detunings and light shifts. Specifically, $\omega_{c}$ is the cavity frequency, $\omega_1$ is the frequency difference between the two active states, $g$ is the single atom-cavity coupling, and $\omega_{r,s}$ is the frequency of the $\sigma_-$- and $\sigma_+$-polarized lasers respectively, with $\Delta_{r,s}$ and $\Omega_{r,s}$ being the related detunings and single atom-laser coupling strengths respectively. By setting $\Delta_{s} = - \Delta_{r}$ we can maximize $U$ and greatly reduce $\omega$. If we choose $g = 20 (2\pi)$MHz, $\Omega_{r,s} = 500 (2\pi)$kHz, $\kappa = 50 (2\pi)$kHz \footnote{See, e.g., \cite{Sames14,Reimann15}, although we choose a smaller $\kappa$ than realized in these experiments. However, we note that good squeezing is still possible with larger $\kappa$.} and $|\Delta_{r,s}| = 3.5 (2\pi)$GHz and assume full tuneability of the small frequency offsets and magnetic field strength that define $\omega$ and $\omega_0$, then we could reach a regime of effective parameters $\set{\omega,\omega_0,U/N,\lambda/\sqrt{N}}/\kappa = \set{0.01,0.01,3.0,0.01}$. While this regime does not offer the idealized squeezing above, it does still offer significant metrological gain over a coherent state.

%%%%%%%%%%%%%%%%%%%%%%%%%%%%%%%%%%%%%%%%%%%%%%%%%%
\subsection{Unresolved Dicke squeezing}

\begin{figure}[b!]
\begin{subfigure}[b]{0.4\textwidth}
\includegraphics[width=\textwidth]{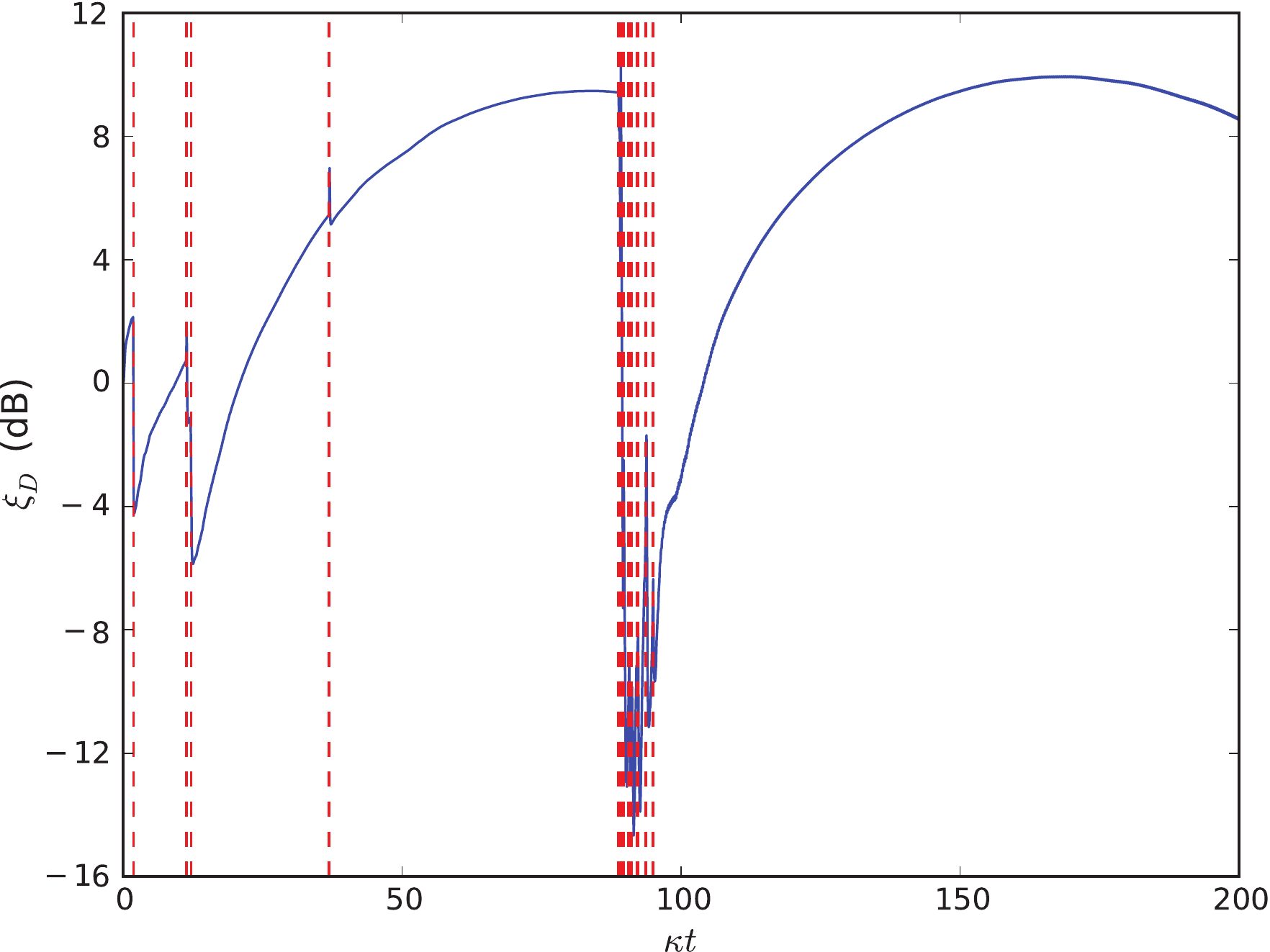}
\caption{}
\end{subfigure}

\begin{subfigure}[b]{.4\textwidth}
\includegraphics[width=\textwidth]{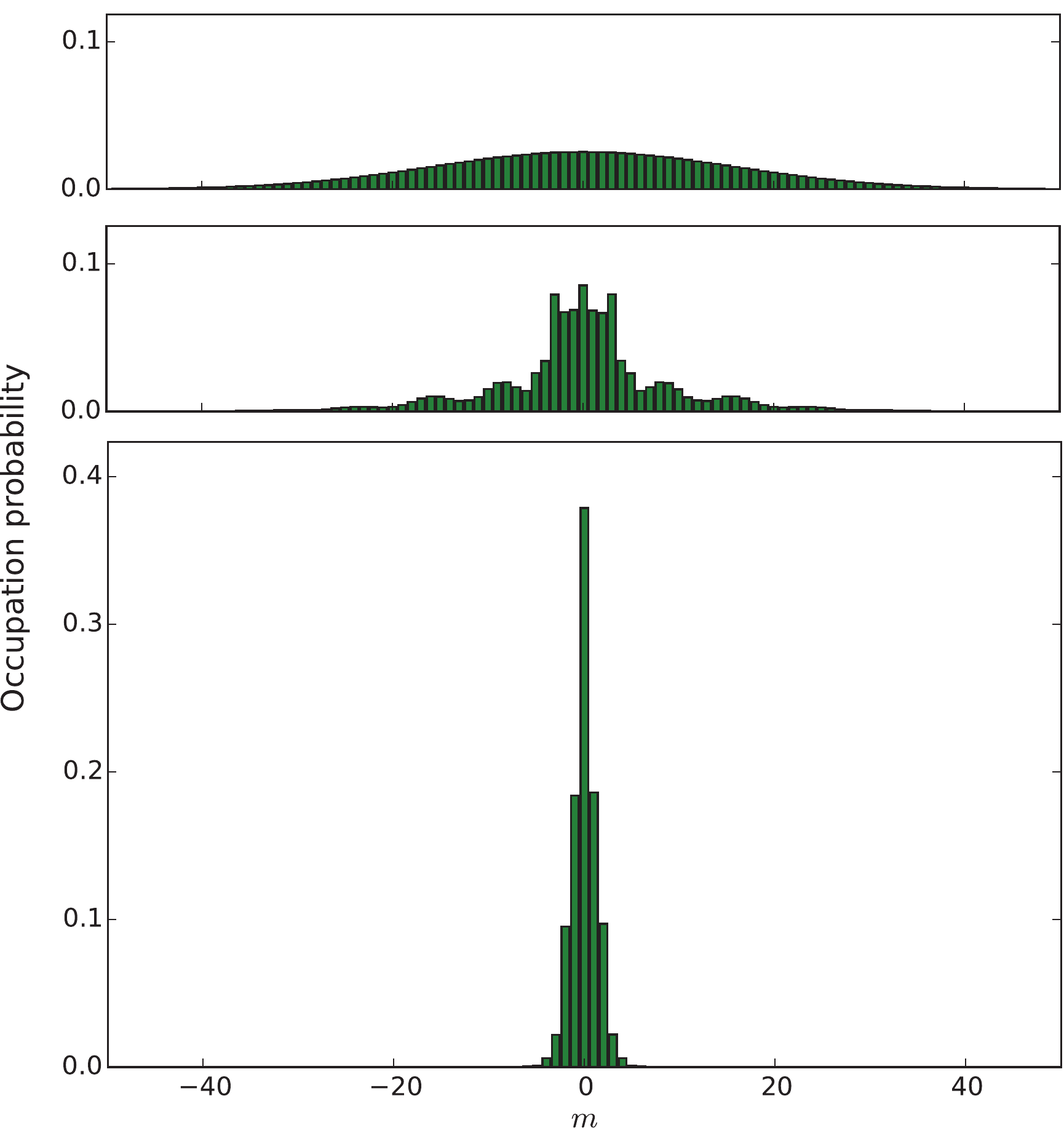}
\caption{}
\end{subfigure}
\caption{For a single trajectory with $N=1000$ and $\set{\omega,\omega_0,U,\lambda}/\kappa = \set{0.01,0.01,3000.0,0.316}$, (a) squeezing with the times of photon detection events represented by red vertical dashed lines and (b) populations initially (top) and after the first jump (middle) and twelfth jump (bottom).\label{tightens}}
\end{figure}

We use these parameters with the atomic ensemble initiated into a coherent spin state centered on the equator of the Bloch sphere. Here, considering single quantum trajectories, we calculate the squeezing \emph{immediately} after each photon detection. As in the ideal case, the probability of a photon arising from population in the central states is higher than from the outer states, but due to the lower value of $U$, the difference is much less sharp. This means that the measurement of a photon tightens the distribution around the central state rather than ``resolving'' a single Dicke state. Due to this, each photon detection shortly afterwards tightens the spread further. In between photon detections, the backaction of the null measurement drains population from the central states and the degree of squeezing worsens. As such, the optimal squeezing is likely to occur when a number of photon detections are measured in a short amount of time. This means that a protocol where one waits for a certain number of photon detections in a certain time frame can achieve substantial Dicke squeezing. It should be noted that such a protocol could also work with high efficiency. For 1000 atoms and the parameters described above, 38.9\% of trajectories run had 12 or more jumps.

%%%%%%%%%%%%%%%%%%%%%%%%%%%%%%%%%%%%%%%%%%%%%%%%%%
%%%%%%%%%%%%%%%%%%%%%%%%%%%%%%%%%%%%%%%%%%%%%%%%%%
\section{Conclusion}

We have shown the existence of strong spin squeezing and entanglement depth in the \emph{steady state} of an open, generalized Dicke model. We have shown that this arises from the collective atomic state being ``pumped'' towards the highly entangled Dicke state $\ket{N/2,0}$, in a manner reminiscent of resolved sideband cooling of atoms trapped in harmonic potentials. By altering the Dicke model parameters, it is also possible to instead pump towards an arbitrary Dicke state $\ket{N/2,m}$. This means that entanglement between every atom in an ensemble could be achieved in steady state, rather than as a transient or probabilistic phenomenon. The steady state nature of the entanglement, even when accessed via a probabilistic method, means that the produced Dicke state is stable. The possibility of producing such a stable entangled state has obvious benefits for quantum information and quantum computing protocols.

While the timescale involved in preparing these states through natural evolution is very large, we have also shown that it is possible to access the steady state on a much shorter timescale via either a single-photon-heralded, probabilistic scheme or by suitable time variation of a parameter of the model. The squeezing is at the Heisenberg limit for the ideal case, but we show that substantial squeezing is still possible with realistic cavity QED parameters.

This work highlights the stark change in dynamics that can occur with the addition of a non-linear term to the Dicke model. The interplay of the linear terms with a simple nonlinear shift of energy levels has been shown here to give rise to exotic steady states. This raises the question of what other possibilities might arise from similar nonlinear terms in, for example, models involving atoms of higher spin \cite{Masson17}. In particular, the spin-1 derivation in  \cite{Masson17} introduces a term of the form $\hat{n}_0 \hat{a}^\dagger \hat{a}$, where $\hat{n}_0$ is the number operator for the $\ket{m=0}$ state of the spin-1 atoms. If this term is allowed to dominate, then we might expect the ensemble to be pumped to states with no atoms in the $\ket{m=0}$ state; i.e., to a family of states containing completely classical states, such as $\ket{m=\pm1}^{\otimes N}$, as well as highly entangled states such as the Dicke state $\ket{N,0}$ with the atoms exactly split between the two states $\ket{m=\pm1}$.

%%%%%%%%%%%%%%%%%%%%%%%%%%%%%%%%%%%%%%%%%%%%%%%%%%
%%%%%%%%%%%%%%%%%%%%%%%%%%%%%%%%%%%%%%%%%%%%%%%%%%

\acknowledgments

The authors acknowledge the contribution of NeSI high-performance computing facilities to the results of this research. New Zealand's national facilities are provided by the New Zealand eScience Infrastructure and funded jointly by NeSI's collaborator institutions and through the Ministry of Business, Innovation and Employment's Research Infrastructure program. The authors also acknowledge support from the Marsden Fund of the Royal Society of New Zealand (Contract No.~UOA1328).

%%%%%%%%%%%%%%%%%%%%%%%%%%%%%%%%%%%%%%%%%%%%%%%%%%
%%%%%%%%%%%%%%%%%%%%%%%%%%%%%%%%%%%%%%%%%%%%%%%%%%


\end{document}